\documentclass[aps,amsmath,twocolumn,superscriptaddress,prb]{revtex4}
\usepackage{amsmath,amssymb,bm}
\usepackage[dvips]{graphicx}
\usepackage{braket}
\usepackage{ulem}
\usepackage{setspace}[10]
\usepackage[dvips]{color}
\usepackage[dvipsnames]{xcolor}

\begin{document}
\title{Magnetic thermal conductivity far above the N\'eel temperatures in the 
Kitaev-magnet candidate $\alpha$-RuCl$_{3}$}

\author{Daichi Hirobe}
\email{daichi.kinken@imr.tohoku.ac.jp}
\affiliation{Institute for Materials Research, Tohoku University, Sendai 980-8577, Japan}

\author{Masahiro Sato}
\email{masahiro.sato.phys@vc.ibaraki.ac.jp}
\affiliation{Spin Quantum Rectification Project, ERATO, Japan Science and Technology Agency, Sendai 980-8577, Japan}
\affiliation{Department of Physics, Ibaraki University, Mito, 
Ibaraki 310-8512, Japan}

\author{Yuki Shiomi}
\affiliation{Institute for Materials Research, Tohoku University, Sendai 980-8577, Japan}
\affiliation{Spin Quantum Rectification Project, ERATO, Japan Science and Technology Agency, Sendai 980-8577, Japan}

\author{Hidekazu Tanaka}
\affiliation{Department of Physics, Tokyo Institute of Technology, 
Meguro-ku, Tokyo 152-8551, Japan}

\author{Eiji Saitoh}
\affiliation{Institute for Materials Research, Tohoku University, Sendai 980-8577, Japan}
\affiliation{Spin Quantum Rectification Project, ERATO, Japan Science and Technology Agency, Sendai 980-8577, Japan}
\affiliation{The Advanced Science Research Center, Japan Atomic Energy Agency, Tokai 319-1195, Japan}
\affiliation{WPI Advanced Institute for Materials Research, Tohoku University, Sendai 980-8577, Japan}

\date{\today}

\begin{abstract}
   We have investigated the longitudinal thermal conductivity of 
$\alpha$-RuCl$_{3}$, the magnetic state of which 
is considered to be proximate to 
a Kitaev honeycomb model, 
along with the spin susceptibility and magnetic
specific heat. 
We found that the temperature dependence 
of the thermal conductivity exhibits an additional peak around 100 K, 
which is well above the phonon peak temperature ($\sim$ 50 K). 
The higher-temperature peak position is comparable 
to the temperature scale of the Kitaev couplings rather than 
the N\'eel temperatures below 15 K.  
The additional heat conduction was 
observed for all five samples used in this study, and 
was found to be rather immune to a structural phase transition of 
$\alpha$-RuCl$_{3}$, which suggests its different origin from phonons. 
Combined with experimental results of the magnetic specific heat, 
our transport measurement suggests strongly that the higher-temperature 
peak in the thermal conductivity is attributed to itinerant spin excitations 
associated with the Kitaev couplings of $\alpha$-RuCl$_{3}$. 
A kinetic approximation of the magnetic thermal conductivity yields 
a mean free path of $\sim$ 20 nm at 100 K, which is well longer than
the nearest Ru-Ru distance ($\sim$ 3 \AA), 
suggesting the long-distance coherent propagation of 
magnetic excitations driven by the Kitaev couplings. 
\end{abstract}

\maketitle
\noindent\textit{Introduction}

Quantum spin liquid is a phase of magnetic insulator in which frustration or 
quantum fluctuation prohibits magnetic order while keeping spin 
correlation~\cite{Balents, Sachdev, XGWen}. It induces rich physical phenomena 
depending on the types of spin liquids, which cannot be realized in standard 
ordered magnets. 
Several quantum-spin models have been proposed as possible realizations of 
quantum spin liquids along with candidate frustrated magnets such as 
$\kappa$-(BEDT-TTF)$_2$Cu$_2$(CN)$_3$ (Refs. \onlinecite{Shimizu, Yamashita2010}), 
ZnCu$_3$(OH)$_6$Cl$_2$ (Ref. \onlinecite{Han}), 
Na$_4$Ir$_3$O$_8$ (Refs. \onlinecite{Okamoto, Fauque})
and $\alpha$-RuCl$_{3}$ (Ref. \onlinecite{Plumb}). 

Among them, the Kitaev honeycomb model is unique in that the ground state is exactly 
calculated and known to be a two-dimensional (2D) quantum spin liquid~\cite{Kitaev}. 
In this model, spin-1/2 moments, $\bm{S}_{\bm{r}}$, sit on 
a honeycomb lattice and interact via the bond-dependent Ising couplings; 
different spin components interact via Ising coupling for the three different 
bonds of the honeycomb lattice. These anisotropic couplings, called the Kitaev couplings 
$J_\alpha S^\alpha_{\bm{r}} S^\alpha_{\bm{r}'}$ ($\alpha=x,y,z$)~(Ref. \onlinecite{Kitaev}), 
yield frustration and induce a quantum spin liquid. 
As a result, the localized spins break down into two fractionalized excitations, 
called itinerant Majorana fermions and gauge fluxes \cite{Kitaev, Baskaran}. 
Several physical quantities of the Kitaev honeycomb model can be captured 
in terms of these fractionalized particles. For example, magnetic entropy release 
was shown to occur at two distinct temperatures using a Monte Carlo simulation \cite{Nasu2015}; 
the higher-temperature release is attributed to the onset of Fermi degeneracy of 
itinerant Majorana fermions, and the lower-temperature one to thermal fluctuations of 
gauge fluxes.

Originally, the Kitaev honeycomb model was introduced as a toy model possessing 
nontrivial topological properties \cite{Kitaev}. However, the model was 
predicted to be realized as a low-energy spin system that 
describes magnetism of spin-orbit coupled Mott insulators~\cite{Jackeli}. 
Moreover, a successive theoretical study showed 
that the spin-liquid ground state survives 
even when a small Heisenberg interaction is added to 
the Kitaev model~\cite{Chaloupka2010}. 
Following these theoretical results, 
actual magnetic compounds have been explored intensively 
\cite{Kim, Singh, Choi, Chun, Sandilands, Sears, Kubota, Majumder, Banerjee, Park, Sandilands2016} 
for the Kitaev quantum spin liquid along with further theoretical studies, 
which unraveled various physical properties of the Kitaev model and 
its variants \cite{Jiang2011,Reuther2011, Chaloupka2013, 
Knolle2014a, Knolle2014b, Yamaji2014, Rau2014, Sato, Nasu2015, 
H-S_Kim, Song2016, Yamaji, Nasu2016}.

   One of the candidate materials is the Mott insulator $\alpha$-RuCl$_{3}$ 
(Refs. \onlinecite{Plumb, Sandilands, Sears, Kubota, Majumder}). In this material, 
Ru$^{3+}$ has an effective spin-1/2 owing to the strong spin-orbit coupling. 
Ru$^{3+}$ ions are located on an almost perfect honeycomb lattice by sharing the edges 
of the octahedral RuCl$_6$ in the $ab$ plane, as shown in Fig. 1a. The Kitaev couplings 
are predicted to arise from superexchange couplings in such a special network of Cl$^-$ ions. 
In fact, the low-temperature magnetic ordering below about 15 K is attributed to proximity 
to the Kitaev spin liquid \cite{Banerjee, Nasu2016}. Even above the ordering temperature, 
properties proximate to the Kitaev spin liquid were reported for $\alpha$-RuCl$_{3}$ 
in recent experiments such as Raman and inelastic neutron scattering measurements 
\cite{Sandilands,Sandilands2016,Banerjee}. From these measurements, 
the strength of the Kitaev couplings was estimated to be about 100 K. In addition, 
measurement of the specific heat \cite{Kubota} suggested magnetic-entropy release at about 90 K, 
which appears consistent with the theoretical prediction for itinerant Majorana fermions 
in the Kitaev spin liquid \cite{Nasu2015, Yamaji}.

\begin{figure}[tth]
\begin{center}
\includegraphics[width=8cm]{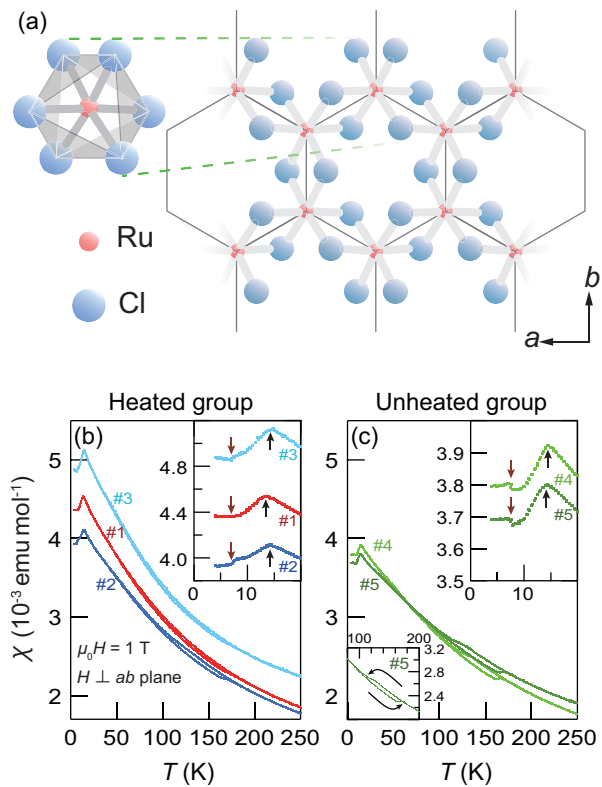}
\end{center}
\caption{(a) Monoclinic crystal structure \textit{C}2/\textit{m} of $\alpha$-RuCl$_3$ 
at room temperature. Red and blue spheres denote Ru$^{3+}$ and Cl$^{-}$ ions, respectively. 
A honeycomb lattice of Ru$^{3+}$ ions is denoted by solid lines, viewed perpendicular to 
the \textit{ab} plane. Shown above to the left is a single RuCl$_6$ octahedron, 
which composes the honeycomb lattice.
(b), (c) Temperature $T$ dependence of the magnetic 
susceptibility $\chi$ for samples \#1-3 (b) and \#4-5 (c). 
A magnetic field, $\mu_0 H$, of 1 T was applied perpendicular 
to the $ab$ plane of $\alpha$-RuCl$_3$.
The insets above to the right in (b) and (c) show a magnified view of 
a low-temperature region. Brown and black arrows indicate N\'eel temperatures of about
8 K and 14 K, respectively. The inset below to the left in (c) shows a magnified view
of a hysteresis loop observed for sample \#5.}
\end{figure}

   Various experimental techniques have been applied to $\alpha$-RuCl$_{3}$; 
however, its transport properties have not been well explored despite the fact 
that transport experiments are often used for detecting signatures of spin liquid 
states and their fractionalized excitations \cite{Sologubenko, Hess, Yamashita2009, Hirobe}. 
In this paper, we experimentally study the thermal conductivity $\kappa$ in $\alpha$-RuCl$_{3}$. 
The temperature dependence of $\kappa$ reveals an anomalous sub-peak that is distinguished 
from a main peak due to phonons. Importantly, the anomalous heat conduction is concomitant 
with the growth of the magnetic specific heat \cite{Kubota} and is maximized at about 100 K; 
the same energy scale of the Kitaev couplings reported for $\alpha$-RuCl$_{3}$. 
Our observations suggest that the sub-peak in $\kappa$ versus $T$ is due to the propagation 
of magnetic excitations driven by the Kitaev couplings.

\noindent\textit{Experimental details}

   Single crystals of $\alpha$-RuCl$_{3}$ were grown by a vertical Bridgman method,
which was exactly the same as reported by one of the present authors \cite{Kubota}.
The grown single crystals were cleaved along the $ab$ plane,
and cut into a cuboid that was typically $5$mm long, $1$mm wide, and $0.5$mm 
thick with the largest surface in the $ab$ plane.
The magnetization of the grown single crystals was measured with a Magnetic Property
Measurement System (Quantum Design, Inc.). 

	The thermal conductivity of $\alpha$-RuCl$_3$ was measured by a standard 
steady-state method \cite{Onose2008, Shiomi2009, Shiomi2010, Onose2010}.
We used five samples for the thermal conductivity
measurement, which were selected from two batches grown separately 
under the same condition:
sample \#1 was selected from one batch while samples \#2-5 from the other.
We used two types of glue to attach the samples to a copper block:
silver-filled epoxy adhesive (H20E, EPO-TEK) for samples \#1-3, which was dried by heating at 
$120\ {}^\circ\mathrm{C}$ for an hour; GE varnish for samples \#4 and \#5, 
which was air-dried without being heated. In the following, samples \#1-3 
are collectively called the heated group while samples \#4-5 the unheated group.
As decribed below, we found that heating $\alpha$-RuCl$_3$ samples modifies 
the magnetic susceptibility, probably attributable 
to crystal stacking faults, but does not affect the thermal conductivity.
A temperature gradient was generated within the $ab$ plane, or along the longitudinal 
direction of the single crystal by using a chip resistance heater (1 kOhm) attached to 
the end of the single crystal. The resulting temperature difference was measured 
using two cernox sensors (CX-1050-BG-HT, 0.75$\times$1.0$\times$0.25 mm$^3$) 
attached to the top surface with 5$\times$1 $\mathrm{mm}$$^2$. 
The magnitude of temperature differences was set to less than five percent of 
the lower temperature of the cernox sensors. The measurement was performed 
in the temperature range between 2.5 K and 300 K under zero magnetic field 
in a high vacuum ($<$ 10$^{-5}$ Torr) using a Physical Property Measurement 
System (Quantum Design, Inc.).

\noindent\textit{Results and discussion}

	We first discuss the effect of heating an $\alpha$-RuCl$_3$
sample on interlayer stacking faults before the thermal conductivity measurement 
(see also the section of experimental details). 
It is known that crystal stacking faults form easily 
along the $c$-axis because of the weak interlayer coupling, 
and affect the structural and magnetic properties of $\alpha$-RuCl$_3$ 
(Refs. \onlinecite{Sears, Kubota, Majumder, Banerjee}).
In Figs. 1(b) and 1(c), we show the temperature $T$ dependence of 
the magnetic susceptibility $\chi$ for a magnetic field perpendicular to the $ab$ plane, 
investigated after the thermal conductivity measurement. 
We found that the behavior of $\chi$ around a N\'eel temperature 
of $\sim8$ K is different between heated and unheated groups 
[see insets to Figs. 1(b) and 1(c)]:
for the heated group, 
$\chi$ decreases smoothly on decreasing $T$ acorss $8$ K 
while it increases rather abruptly for the unheated group.
Additionally, we found another change 
in a much higher-$T$ region. 
For the unheated group [Fig. 1(c)], a hysteresis loop with a width of $\sim10$ K 
is clearly resolved between 120 K and 165 K, which arises from 
a structural phase transition via the strong spin-orbit coupling \cite{Park}. 
In contrast, such a clear loop disappears for the heated group, 
replaced with the much smaller and broader loop [Fig. 1(b)].
Such sample dependence of the magnetic properties has been reported, 
and attributed to the different proportion of crystal stacking faults \cite{Banerjee}. 
In the present study, the structural phase transition is blurred in the heated group,
probably by a larger region of crystal stacking faults introduced under the heating
procedure.

Having observed different hysteresis loops of the structural phase transition 
between the heated and unheated groups, we now turn to thermal conductivity results.
   Figure 2(a) shows the $T$ dependence of $\kappa$ of the heated group. 
$\kappa$ increases monotonically with decreasing $T$ from 300 K to 180 K. 
This is attributed to the growing mean free path of phonons; the rate of 
Umklapp scattering decreases with decreasing $T$ in this temperature 
range \cite{Ziman}. With decreasing $T$ further, however, 
the $T$ dependence of $\kappa$ deviates from a known function of
the phonon thermal conductivity, and $\kappa$ begins increasing more rapidly 
from 180 K. As a result, a sub-peak structure forms around 110 K, 
which is reminiscent of the spin thermal conductivity in low-dimensional 
quantum spin systems \cite{Sologubenko, Hess, Kawamata, Hlubek, Nakamura}.
$\kappa$ then reaches a main peak at $\sim$ 50 K and decreases 
strongly upon decreasing $T$ further. The main peak is simply 
due to phononic heat conduction, the position of which 
($\sim50$ K) results from competition between the increasing 
mean free path and the decreasing number of phonons with decreasing 
$T$ (Refs. \onlinecite{Ziman, Ashcroft}). 

Crucially, the thermal conductivity $\kappa$ of the unheated group, 
shown in Fig. 2(b), exhibits a sub-peak in the same position 
as that for the heated group despite the clear hysteresis loop 
in the magnetic susceptibility data [see also Fig. 1(b)]. 
The result indicates a low correlation
of $\kappa$ with the structural phase transition. 
   This is also supported by comparing $\kappa$ 
measured while increasing and decreasing $T$. The inset to Fig. 2(a) 
compares $\kappa$ around 150 K measured in such scans of $T$.
A hysteresis loop is not observed over the 
sub-peak of $\kappa$ within the error range of 0.2 W\hspace{0.3ex}K$^{-1}\hspace{0.3ex}$m$^{-1}$ 
for sample \#1 and 0.1 W\hspace{0.3ex}K$^{-1}\hspace{0.3ex}$m$^{-1}$
for samples \#2-5. The observation shows that the sub-peak is insensitive 
to the structural phase transition, which is consistent with the same $T$
dependence of the sub-peak shared by the heated and unheated groups.
This means that a possible change in the phonon thermal conductivity
caused by the structural phase transition is too small to be 
observed. Accordingly, the phonon thermal conductivity may be 
approximated by the conventional one that exhibits a single peak
as a function of $T$ in a practical analysis; a different origin 
from phonons is needed to explain the sub-peak structure.
We note that a sharp peak of the phonon thermal conductivity
begins growing from 8 K [see also the inset to Fig. 2(b)], 
which is comparable with a N\'eel temperature shown in Fig. 1. 
This increase in the phonon thermal conductivity may originate 
from the suppression of phonon scattering off paramagnetic 
fluctuations \cite{Sharma}.

\begin{figure}[tth]
\begin{center}
\includegraphics[width=6cm]{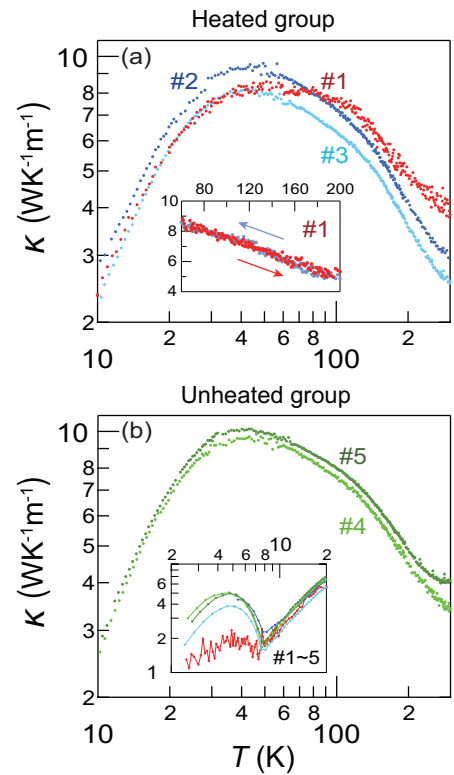}
\end{center}
\caption{(a), (b) Temperature \textit{T} dependence of 
the thermal conductivity $\kappa$ for samples \#1-3 (a) 
and \#4-5 (b). The inset to (a)
shows a magnified view of a temperature region for sample \#1 
in which the structural phase transition occurs. Red and 
sky-blue data points were taken while increasing and decreasing \textit{T}, 
respectively. The inset to (b) shows a magnified view of 
a low-temperature region for samples \#1-5.}
\end{figure}


   We extract the sub-peak component of $\kappa$ by assuming 
the 2D Debye model for the phonon component. Using an elementary kinetic theory 
\cite{Ziman, Ashcroft}, the phonon thermal conductivity is given by 
$\kappa_{\mathrm{ph}}=C_{\mathrm{ph}}\, v_{\mathrm{ph}}\, l_{\mathrm{ph}}/2$, 
where $C_{\mathrm{ph}}$ is the lattice specific heat, $v_{\mathrm{ph}}$ is 
the velocity, and $l_{\mathrm{ph}}$ is the mean free path of the phonons. 
For $C_{\mathrm{ph}}$, we referred to the procedure in Ref. \onlinecite{Kubota}; 
the Debye temperature calculated thereby for $\alpha$-RuCl$_{3}$ was 257 K 
and $v_{\mathrm{ph}}$ was calculated to be 1,490 m\hspace{0.3ex}s$^{-1}$ 
using this Debye temperature. For $l_{\mathrm{ph}}$, we used an empirical formula, 
$l^{-1}_{\mathrm{ph}}=L^{-1}+A\, T^n\, \mathrm{e}^{-T_0/T}$, with $L$, $A$, $n$, $T_0$ 
being fitting parameters \cite{Sologubenko, Kawamata}. Using these parameters, 
we successfully fitted $\kappa_{\mathrm{ph}}$ to  the total 
$\kappa$ in the range of 20 K $<$ $T$ $<$ 60 K plus 200 K $<$ $T$ $<$ 270 K, 
so that $\kappa_{\mathrm{ph}}$ reproduced the phonon peak at ~50 K and was asymptotic 
to $\kappa$ towards high temperature [see also Fig. 3(a) and Ref. \onlinecite{fittingPhonon} 
for a result of fitting].

The difference $\Delta\kappa = \kappa - \kappa_\mathrm{ph}$ is plotted 
as a function of $T$ in Fig. 3(b). A broad peak reflecting the sub-peak in Fig. 2 
appears also in $\Delta\kappa$ and reaches a maximum of 
$\sim$ 0.5 W\hspace{0.3ex}K$^{-1}$\hspace{0.3ex}m$^{-1}$ at about $T_\mathrm{p} = 110$ K. 
Remarkably, the broad peak is concomitant with a broad maximum reported for the $T$ dependence 
of the magnetic specific heat $C_\mathrm{mag}$ of an $\alpha$-RuCl$_{3}$ sample grown 
by the same method \cite{Kubota}, as shown in Fig. 3(c). Since the broad maximum of 
$C_\mathrm{mag}$ without magnetic phase transitions is qualitatively consistent with 
a theoretical calculation based on itinerant Majorana fermions \cite{Nasu2015, Yamaji}, 
the agreement between $\Delta\kappa$ and $C_\mathrm{mag}$ suggests that 
the itinerant quasiparticles carry heat. 
It should be noted that the agreement is confirmed 
for both heated and unheated groups along with the similar magnitude of $\Delta\kappa$,
despite the different $T$ dependence of $\chi$ between these two groups.
This suggests that the itinerant quasiparticles are rather immune to 
crystal stacking faults, and to the resulting variation in interlayer 
magnetic couplings relevant to three-dimensional N\'eel orders; 
the dynamics of the quasiparticles are expected to be governed by 
2D magnetic couplings that are much stronger than the interlayer ones.
We also emphasize that conventional spin waves in magnetically ordered states are 
irrelevant to the peak around $T_\mathrm{p} = 110$ K as $T_\mathrm{p}$ 
is much higher than the ordering temperatures ($<$ 15 K) of $\alpha$-RuCl$_{3}$. 
Instead, $T_\mathrm{p}$ corresponds to the strength of the 2D Kitaev couplings 
reported for $\alpha$-RuCl$_{3}$ ($\sim$ 100 K). In this high-$T$ range, 
a continuum of Raman scattering \cite{Sandilands}, 
persistent short-range spin correlation \cite{Banerjee, Sandilands2016} 
as well as fermionic response \cite{Nasu2016} were reported and attributed 
to 2D Kitaev coupling. 

\begin{figure}[tth]
\begin{center}
\includegraphics[width=8.5cm]{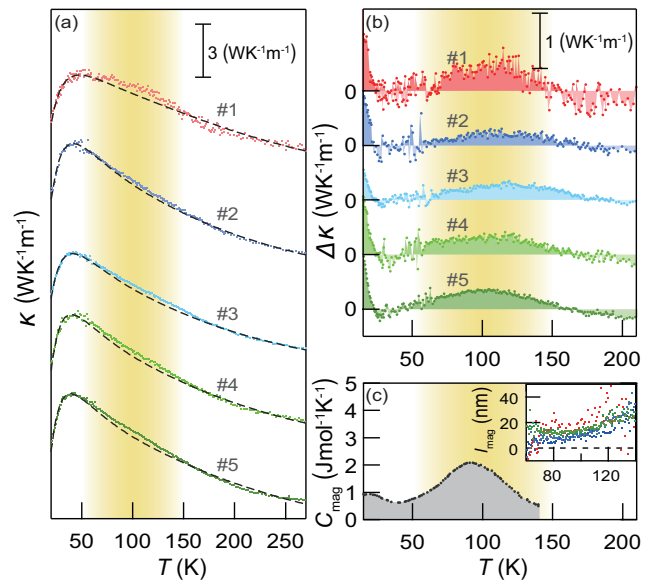}
\end{center}
\caption{ (a) Temperature $T$ dependence of the total thermal conductivity $\kappa$
(circles) presented in Fig. 2 and the phonon thermal conductivity $\kappa_\mathrm{ph}$ 
(broken lines) estimated using the Debye model (see text). 
(b) \textit{T} dependence of a deviation of $\kappa$ from $\kappa_\mathrm{ph}$, 
$\Delta\kappa = \kappa - \kappa_\mathrm{ph}$.
(c) $T$ dependence of the magnetic specific heat $C_\mathrm{mag}$ of 
$\alpha$-RuCl$_3$. $C_\mathrm{mag}$ was estimated by subtracting the lattice 
specific heat from the specific heat of $\alpha$-RuCl$_3$, 
following the procedure in Ref. \onlinecite{Kubota}. 
The inset shows the $T$ dependence of the the mean free path $l_\mathrm{mag}$  
of magnetic excitations in $\alpha$-RuCl$_3$ estimated using an elementary kinetic
theory (see text). 
Data of the thermal conductivity and the specific heat were taken using 
$\alpha$-RuCl$_3$ samples grown by the same group and the same method.}
\end{figure}

We are in a position to characterize $\Delta\kappa$ presumably related 
with the Kitaev couplings of $\alpha$-RuCl$_{3}$. 
Below, we assume that a paramagnetic state of 
$\alpha$-RuCl$_{3}$ is proximate to the Kitaev spin liquid and 
that the broad peak in $\Delta\kappa$ originates from 
itinerant Majorana fermions. 
The assumption allows us to use
the 2D elementary kinetic theory for the magnetic thermal conductivity 
$\kappa_\mathrm{mag}$, given by 
$\kappa_\mathrm{mag}=C_\mathrm{mag}\, v_\mathrm{mag}\, l_\mathrm{mag}/2$ 
to analyze $\Delta\kappa$. 
Here, $v_\mathrm{mag}$ and $l_\mathrm{mag}$ are, respectively, 
the velocity and the mean free path of the magnetic excitations. 
We note that the quasi-particle picture is supported also 
by a Fermi-liquid behavior identified in the numerical calculation 
in Ref.~\onlinecite{Nasu2015}.
Microscopic models have not been established yet to quantitatively capture the magnetic properties 
of $\alpha$-RuCl$_{3}$; to estimate $l_\mathrm{mag}$, therefore, we simply 
assume that $\kappa_\mathrm{mag}$ is driven by the Kitaev couplings 
$J^\alpha\sim$ 100 K ($\alpha=x,y,z$) alone and that $v_\mathrm{mag}$ 
is given by averaging the Majorana-fermion group 
velocity of the Kitaev honeycomb model~\cite{Kitaev, Sato, Nussinov} 
in a certain region of the Brillouin zone~\cite{groupVelocity}. 
The velocity $v_\mathrm{mag}$ is 
then found to be 
1,620 m\hspace{0.3ex}s$^{-1}$ 
by setting the Kitaev couplings to be 100 K.
By putting $\Delta\kappa = \kappa_\mathrm{mag}$ and using the experimental data of 
$C_\mathrm{mag}$ and $\kappa_\mathrm{mag}$, $l_\mathrm{mag}$ was found to be $10\sim20$ nm 
between 60 K and 110 K \cite{compare1Dspinons}, into which the data for all the samples 
fit as shown in the inset to Fig. 3(c).
The result implies that the excitations can carry entropy over a distance up to 60 times 
as long as the nearest Ru-Ru distance ($\sim$ 3 \AA). 
Although our estimation is reliable only in a qualitative level, 
it suggests that the anomalous heat conduction is due to the coherent 
propagation of itinerant spin excitations 
around $T_\mathrm{p}$, which has not been revealed by measurements of 
the magnetic susceptibility nor the specific heat performed so far. 
Further evidence for the magnetic origin may be derived from investigating
the purely phononic heat conduction in a nonmagnetic analogue of $\alpha$-RuCl$_{3}$,
for example ScCl$_{3}$, that has a similar honeycomb lattice \cite{ScCl3sample}.

\noindent\textit{Conclusion}

In this study, we have investigated the temperature dependence of the thermal 
conductivity of the Mott insulator $\alpha$-RuCl$_{3}$, 
the magnetism of which is related with a Kitaev honeycomb model. 
The thermal conductivity was observed to exhibit a sub-peak structure around 110 K 
that is insensitive to interlayer crystal stacking faults.
Compared with the temperature dependence of 
the magnetic specific heat, the broad peak 
in the thermal conductivity 
was attributed to itinerant spin excitations. Applying a kinetic approximation to 
the magnetic thermal conductivity yielded a long mean free path that was 
up to 60 times longer than the nearest inter-spin distance between 
60 K and 110 K. This result suggests the coherent propagation of itinerant spin excitations 
due to the Kitaev couplings, possibly itinerant Majorana fermions. 

\noindent\textit{Note added.} 

After submitting our original manuscript, 
we became aware of experimental work by Ian A. Leahy \textit{et al.} \cite{Leahy} and
Richard Hentrich \textit{et al.} \cite{Hentrich}, who both focused on the high-field 
dependence of the thermal conductivity of $\alpha$-RuCl$_3$ at low temperatures. 
Their results obtained at zero field are well consistent with ours, 
especially a kink around 8 K. In addition, Seung-Hwan Do \textit{et al.} \cite{Do} 
recently reported the temperature dependence of the magnetic specific 
heat of $\alpha$-RuCl$_3$, and observed a broad peak structure at 
about 100 K, consistent with our result. We also became aware of 
theoretical work by Joji Nasu \textit{et al.} \cite{Nasu2017}. 
Their computation with the use of the Kitaev model showed that 
itinerant Majorana fermions contribute to the longitudinal thermal conductivity 
and the magnetic specific heat within the same temperature range, 
which qualitatively reproduces our result.

\noindent\textit{Acknowledgements}

This work is supported by JSPS (KAKENHI No. JP16H00977, No. 16K13827 and 
No. 26247058 and the Core-to-Core program ``International research center for new-concept 
spintronics devices") and MEXT (Innovative Area ``Nano Spin Conversion Science" (No. 26103005)). 
D. H. is supported by the Yoshida Scholarship Foundation through the Doctor 21 program.
M. S. was supported by Grant-in-Aid for Scientific Research on
Innovative Area, ”Nano Spin Conversion Science” (Grant No.17H05174), 
and JSPS KAKENHI (Grant No. 17K05513 and No. 15H02117).


\begin{thebibliography}{12}
\bibitem{fittingPhonon}
The best fitting to the data points was done in the following range 
of the parameters: 
$1.8\times10^{-7}$ m $<L<2.0\times10^{-7}$ m; 
$3.1\times10^{-5}$ m$^{-1}$ $<A <10.2\times10^{-5}$ m$^{-1}$; 
$1.0<n<1.3$; $31.5<T_0<51.0$.
\bibitem{groupVelocity}
To be specific, $v_\mathrm{mag}$ was calculated by averaging 
the Majorana-fermion group velocity along the 
$\varGamma$-$K$ line, 
using the isotropic Kitaev model 
($J^x=J^y=J^z=J$) at $T=0$:
$v_\mathrm{mag}=(\varDelta k)^{-1}\left|\int^{\varDelta k}_0\mathrm{d}k
\frac{1}{\hbar}\frac{\partial{E}}{\partial k}\right|=\frac{1}{\hbar}\frac{3}{2}\frac{J}{\varDelta k}$. 
Here, $\varDelta k$ denotes the distance between the $\varGamma$ point 
and a $K$ point in wavenumber space.
\bibitem{compare1Dspinons}
For comparison, we note that the mean free path of one-dimensional spinons 
in Sr$_2$CuO$_3$ reaches 50 to 100 nm at 100 K, depending 
on the purity of the primary chemicals
\cite{Sologubenko, Kawamata, Hlubek}.
\bibitem{ScCl3sample}
We tried to measure the thermal conductivity of single crystalline 
ScCl$_3$ in the present study; however, ScCl$_{3}$ samples were found to 
chemically deteriorate during the preparation for thermal conductivity 
measurement, caused by the high dissolubility in water in air. Fixing this problem 
and performing a control experiment will be a future task of importance.
\end{thebibliography}
\end{document}